\documentclass{MolSpaLab}
\usepackage{graphicx}
\usepackage{natbib}
\bibpunct[]{(}{)}{;}{a}{}{,}

\def\apjl{ApJL}
\def\apj{ApJ}

\def\apjs{ApJS}

\begin{document}
\TitreGlobal{Molecules in Space \& Laboratory}
\title{The search for complex molecules in the ISM:\\
       a complete 3~mm line survey of Sgr B2-N and -M}
\author{\FirstName A. \LastName Belloche}
\address{Max-Planck Institut f{\"u}r Radioastronomie, Auf dem H{\"u}gel 69, 
  D-53121 Bonn, Germany}
\author{\FirstName C. \LastName Comito$^1$}
\author{\FirstName C. \LastName Hieret$^1$}
\author{\FirstName K.~M. \LastName Menten$^1$}
\author{\FirstName H.~S.~P. \LastName M{\"u}ller$^{1,}$}
\address{I. Physikalisches Institut, Universit{\"a}t zu K{\"o}ln,
             Z{\"u}lpicher Str. 77, D-50937 K{\"o}ln, Germany}
\author{\FirstName P. \LastName Schilke$^1$}

\runningtitle{Complete 3~mm line survey of Sgr B2-N and -M}
\setcounter{page}{1}

\maketitle
\begin{abstract}

Famous for the extraordinary richness of its molecular content, the Sgr B2
molecular cloud complex is the prime target in the long-standing search for
ever more complex species. We have completed a molecular line survey of the
hot dense cores Sgr B2(N) and SgrB2(M) in the 3~mm wavelength range with the 
IRAM 30\,m telescope. We performed the analysis of this huge data set by 
modeling the whole spectrum at once in the LTE approximation.
Ongoing analyses yield an average line density of about 100 features/GHz
above 3$\sigma$ for Sgr B2(N), emitted and/or absorbed by
a total of 51 molecular species. We find lines from 60
rare isotopologues and from 41 vibrationally excited states in
addition to the main species, vibrational ground state lines. For
Sgr B2(M), we find about 25 features/GHz above 3$\sigma$, from 41 molecular 
species plus 50 isotopologues and 20 vibrationally excited states. 
Thanks to the constant updates to the Cologne Database for Molecular 
Spectroscopy, we are working our way through the assignment of the 
unidentified features, currently 40$\%$ and 50$\%$ above 3$\sigma$ for 
Sgr B2(N) and SgrB2(M), respectively.

\end{abstract}
%
\section{Introduction}
\label{s:intro}

With several active regions and a total
mass of more than $10^6$ M$_\odot$,  Sagittarius B2 (hereafter SgrB2) is one 
of the most complex and massive sites of star formation in the Galaxy.
It is located close to the galactic center and harbors two hot dense cores (M 
and N), which fragment into several sub cores. SgrB2(N) has a very rich 
chemistry and was called the Large Molecule Heimat (LMH) by \citet{Snyder94}. 
Many complex molecules were discovered there like, e.g., acetic acid 
\citep[CH$_3$COOH,][]{Mehringer97}, glycolaldehyde 
\citep[CH$_2$(OH)CHO,][]{Hollis00}, and acetamide 
\citep[CH$_3$CONH$_2$,][]{Hollis06}. It is therefore a prime target to look for
new complex molecules.

Line surveys at (sub)mm wavelengths are needed to search for large complex 
molecules since these molecules emit hundreds of (weak) lines. Regions with 
rich chemistry produce very crowded spectra with many blended lines and the 
confusion limit is easily reached. In this context, the secure identification 
of a new molecule requires the identification of many lines and no single line
predicted by a temperature and column density derived from multiple lines
should be missing \citep[see, e.g., the rebuttal of glycine by][]{Snyder05}. 
Since the spectra are complex, modeling the emission of all known molecules 
is necessary to prevent mis-assignments and point out blended lines before 
claiming the detection of a new molecule.

To search for new complex molecules, we carried out a complete line survey at
3~mm toward the two hot cores SgrB2(N) and SgrB2(M) and we present here some 
preliminary results.


\begin{figure}[t]
\begin{center}
\includegraphics[width=9.0cm,angle=270]{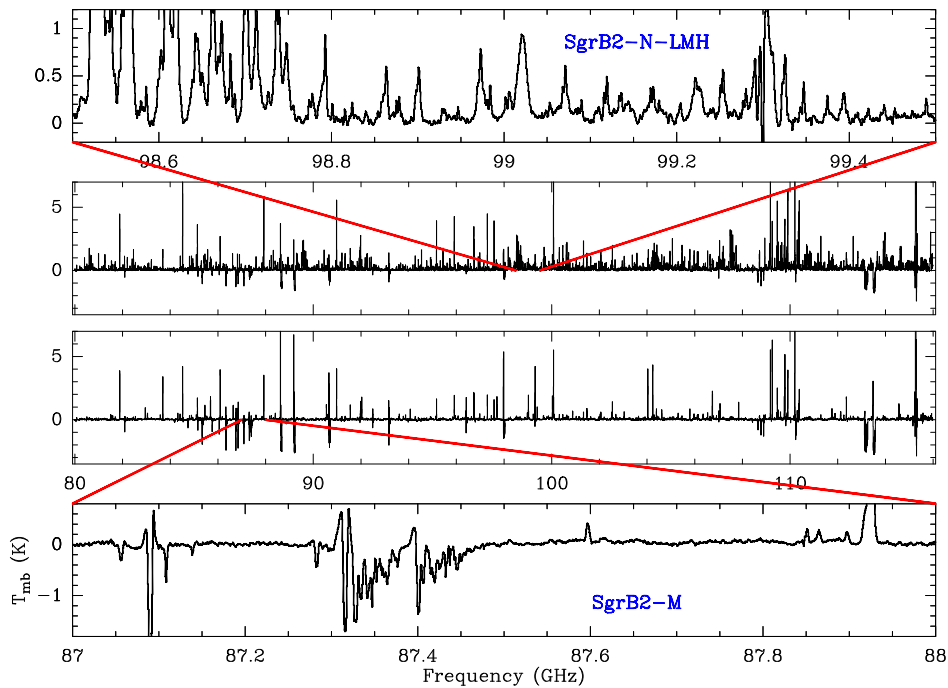}
\caption{Spectra obtained toward SgrB2(N) (\textit{top}) and SgrB2(M) 
(\textit{bottom}) with the IRAM 30\,m telescope between 80 and 116 GHz. The 
blow-ups reveal the full complexity of the SgrB2(N) spectrum and the 
prominent multi-component absorption lines in the SgrB2(M) spectrum .}
\label{f:survey}
\end{center}
\end{figure}

\section{Observations}
\label{s:obs}

We carried out a complete line survey of SgrB2(N) and SgrB2(M) between 80 and 
116 GHz with the IRAM 30\,m telescope in 2004 and 2005, as well as partial 
surveys 
at 2~mm and 1.3~mm. We obtained an rms sensitivity of 15-30 mK at 3~mm in 
$T_a^\star$ scale and reached the confusion limit at 1.3~mm. The full 3~mm 
spectra are shown in Fig.~\ref{f:survey}, as well as a closer view on two 
small frequency ranges. The average line density above 3$\sigma$ is about 100
and 25 features per GHz, translating into about 3700 and 950 lines over the
whole 80-116 GHz band, for SgrB2(N) and SgrB2(M), respectively. Some lines are
seen in absorption with numerous velocity components and are produced by the
diffuse spiral arm clouds lying along the line of sight (Hieret et al.,
\textit{in prep.}).

\section{Modeling and preliminary results}
\label{s:results}

\begin{figure}[t]
\begin{center}
\includegraphics[width=9.0cm,angle=270]{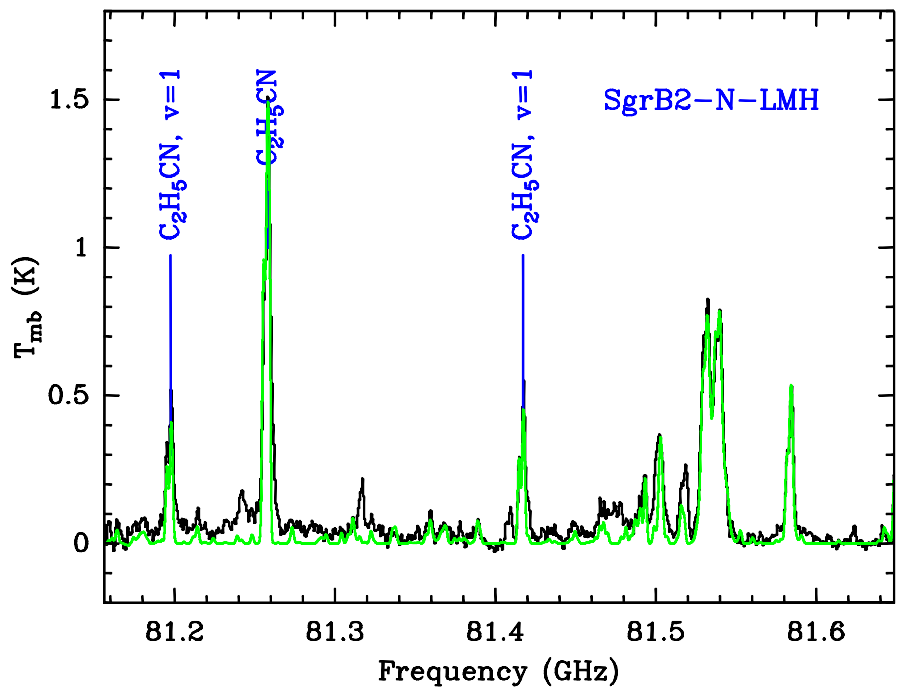}
\caption{Sample spectrum toward SgrB2(N) observed with the IRAM 30\,m telescope.
The green spectrum represents our LTE model including all the molecules we 
have identified so far. A few line assignements are labeled in blue 
(ethylcyanide in the ground and first vibrationally excited states).}
\label{f:model}
\end{center}
\end{figure}

We model the emission of all known molecules, including vibrationally and 
torsionally excited states, and their isotopologues. We use the XCLASS 
software \citep[see][, and references therein]{Comito05} to model the emission
and absorption lines in the LTE approximation.  These
calculations take into account beam dilution, lines opacity, and
blending. The molecular spectroscopic parameters are taken from our line
catalog which mainly contains all entries from the Cologne 
Database for Molecular Spectroscopy \citep[CDMS, see][]{Mueller05} and from 
the molecular spectroscopic database of the Jet 
Propulsion Laboratory \citep[JPL, see][]{Pickett98}. Each molecule is modeled
separately. The whole spectrum including all the identified molecules is then
computed at once (see the example in Fig.~\ref{f:model}), and the parameters 
for each molecule are adjusted again when necessary. The quality of the fit is 
checked by eye over the whole frequency coverage of the line survey. 

The detailed results of this modeling will be published in a forthcoming 
article (Belloche et al., \textit{in prep}). So far, we have identified 51 
molecules, 60 isotopologues, and 41 vibrationally/torsionally excited
states in SgrB2(N), which represent about 60$\%$ of the lines detected above
3$\sigma$. In SgrB2(M), the corresponding numbers are 41, 50, 20, and 50$\%$, 
respectively. We detected very few deuterated species, and only simple ones.

Currently, there are still 40$\%/$50$\%$ lines unidentified in our survey of
SgrB2(N) and SgrB2(M), respectively, and 15$\%/$7$\%$ are even stronger than 
0.3 K (i.e. $\sim 10$ times our noise level). Although most of these 
unidentified lines are most likely emitted by 
vibrationally or torsionally excited states of already known molecules
missing in our database (e.g. ethanol C$_2$H$_5$OH $v_x$=1, vinyl cyanide 
C$_2$H$_3$CN $v_{11} = 2/v_{15}=1$,  ethyl cyanide C$_2$H$_5$CN v=2), there 
is room for detection of new complex molecules. In particular, we have strong
evidence that we detected the molecule aminoacetonitrile -- a likely 
precursor of the simplest amino acid, glycine -- in our 30\,m survey. This 
detection is supported by follow-up observations we carried out with the 
Plateau de Bure Interferometer and the Australia Telescope Compact Array in
2006 (Belloche et al., \textit{to be submitted}). This would be a step forward 
in the long standing search for amino acids.



\end{document}